\documentclass[twocolumn,showpacs,aps,prx,superscriptaddress,longbibliography]{revtex4-2}

\usepackage{color}
\usepackage{hyperref}
\usepackage{graphicx}
\usepackage{amsmath}
\usepackage{mathtools}
\usepackage{color}
\definecolor{darkgreen}{rgb}{0,0.6,0}
\definecolor{orange}{rgb}{0.99,0.257,0}
\usepackage{hyperref}

\usepackage{graphicx}
\usepackage{amsmath}
\usepackage{mathtools}
\usepackage{xr}
\usepackage{soul}
\begin{document}

\title{Environmental memory facilitates search with home return}
\author{Amy Altshuler}
\affiliation{The Raymond and Beverley School of Chemistry, Tel Aviv University, Tel Aviv 6997801, Israel.}
\affiliation{Center for Physics and Chemistry of Living Systems, Tel Aviv University, Tel Aviv, 6997801, Israel.}
\author{Ofek Lauber Bonomo}
\affiliation{The Raymond and Beverley School of Chemistry, Tel Aviv University, Tel Aviv 6997801, Israel.}
\affiliation{Center for Physics and Chemistry of Living Systems, Tel Aviv University, Tel Aviv, 6997801, Israel.}
\author{Nicole Gorohovsky}
\affiliation{Department of Materials Science and Engineering, Tel Aviv University, Tel Aviv, 6997801, Israel.}
\author{Shany Marchini}
\affiliation{Department of Materials Science and Engineering, Tel Aviv University, Tel Aviv, 6997801, Israel.}
\author{Eran Rosen}
\affiliation{The Raymond and Beverley School of Chemistry, Tel Aviv University, Tel Aviv 6997801, Israel.}
\author{Ofir Tal-Friedman}
\affiliation{The Raymond and Beverley School of Physics \& Astronomy, Tel Aviv University, Tel Aviv 6997801, Israel.}
\author{Shlomi Reuveni} 
\thanks{Equal contribution}\email{shlomire@tauex.tau.ac.il}
\affiliation{The Raymond and Beverley School of Chemistry, Tel Aviv University, Tel Aviv 6997801, Israel.}
\affiliation{Center for Physics and Chemistry of Living Systems, Tel Aviv University, Tel Aviv, 6997801, Israel.}
\author{Yael Roichman}\thanks{Equal contribution}
\email{roichman@tauex.tau.ac.il}
\affiliation{The Raymond and Beverley School of Chemistry, Tel Aviv University, Tel Aviv 6997801, Israel.}
\affiliation{Center for Physics and Chemistry of Living Systems, Tel Aviv University, Tel Aviv, 6997801, Israel.}
\affiliation{The Raymond and Beverley School of Physics \& Astronomy, Tel Aviv University, Tel Aviv 6997801, Israel.}

\begin{abstract}
    Search processes in the natural world are often punctuated by home returns that reset the position of foraging animals, birds, and insects. Many theoretical, numerical, and experimental studies have now demonstrated that this strategy can drastically facilitate search, which could explain its prevalence. To further facilitate search, foragers  also work as a group: modifying their surroundings in highly sophisticated ways e.g., by leaving chemical scent trails that imprint the memory of previous excursions. Here, we design a controlled experiment to show that the benefit coming from such ``environmental memory'' is significant even for a single, non-intelligent, searcher that is limited to simple physical interactions with its surroundings. To this end, we employ a self-propelled bristle robot that moves randomly within an arena filled with obstacles that the robot can push around. To mimic home returns, we reset the bristle robot's position at constant time intervals. We show that trails created by the robot give rise to a form of environmental memory that facilitates search by increasing the effective diffusion coefficient. Numerical  simulations, and theoretical estimates, designed to capture the essential physics of the experiment support our conclusions and indicate that these are not limited to the particular system studied herein.
\end{abstract}

\maketitle
\subsection*{Introduction}\label{sec1}


Search processes are common in nature, from animal forging on the macroscopic scale \cite{Physics_of_Foraging,kagan2015search} to the search of bio-molecules inside living cells \cite{von1989facilitated,hammar2012lac}. Over the years, search and first-passage problems attracted considerable attention in different fields and contexts \cite{redner_guide_2001,Metzler_FP,bray2013persistence}. It has been widely observed that a proper choice of the stochastic mode of motion can significantly facilitate search. Finding optimal search strategies in different conditions and under various constraints has thus become a central goal of search research \cite{viswanathan1999optimizing,shlesinger_search_2006,eliazar2007searching,lomholt2008levy,search_strategies,volpe_topography_2017,bernardi2022run}.

Recently, a series of theoretical, computational, and experimental studies demonstrated that resetting a search process repeatedly can accelerate it regardless of the underlying mode of stochastic motion \cite{evans2011,kusmierz2014first, FPTsholomi, chechkin_random_2018, tal-friedman_experimental_2020, Evans2020, blumer2022stochastic, yin2023restart}, and as long as the first-passage time distribution without resetting is sufficiently dispersed \cite{pal2022inspection}.  This counter-intuitive fact was pointed out as a major advantage of search with home returns \cite{pal2020search} that is prevalently displayed by foraging animals, birds, and insects. 


\begin{figure*} [th]
    \centering
    \includegraphics[width=\textwidth]{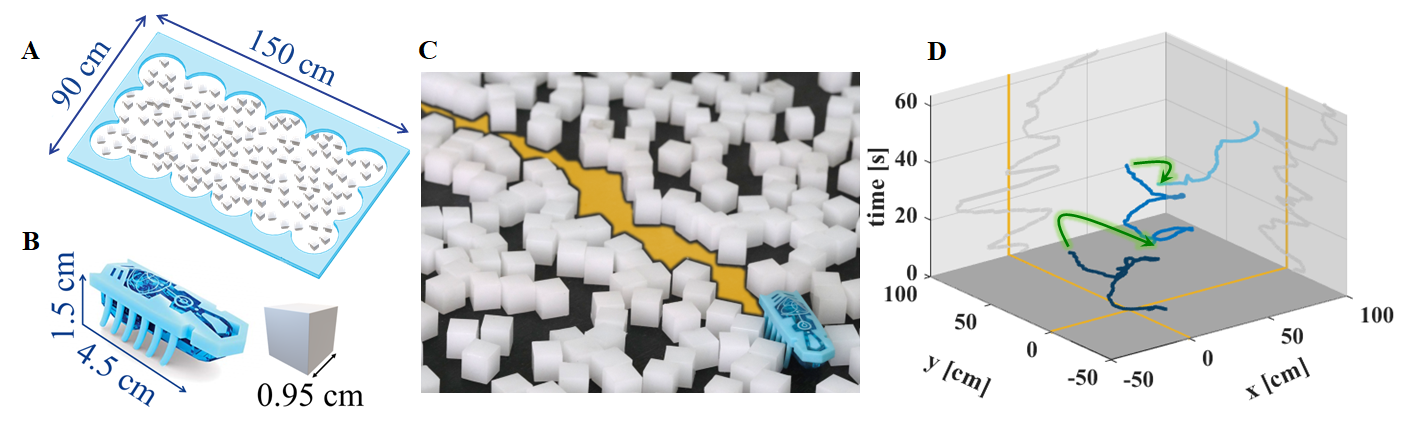}
        \caption{ The experimental setup. \textbf{A}, A sketch detailing the arena and its dimensions (the inner dimensions are $137\times63$~$cm^2$). \textbf{B}, Characteristic scales of the bbot and the mobile cubic obstacles. \textbf{C}, An image of a bbot (in blue) paving its way through the arena and leaving a trail (orange). \textbf{D}, Sample trajectory of the bbot (in shades of blue) and its projections onto the axes (in gray). Resetting is conducted at constant time intervals of $\tau=20$~$seconds$. Resetting events are marked with green arrows.}
    \label{fig:setup}
\end{figure*}

Missing completely from the discussion and analysis of  resetting, and search with home returns, are interactions between the searching agent and its environment. For example, auto-chemotactic cells mark and sense their environment \cite{ben2000cooperative, kranz2016effective, taktikos2011modeling}, thus building memory which can in turn be used to expedite search \cite{meyer2021optimal}. Similarly, ants (and other insects) mark the ground with chemical scents while foraging and navigating back and forth from the nest in search of a food source \cite{holldobler1990ants,david2009trail}.  Here again, memory is gradually built into the environment, which means that the first search attempt is different from the second, that is in turn different from the third, and so on and so forth. 

Ants tend to follow dominant scent trails that were previously used and marked by many ants, indicating that scent trails provide effective means of communication. This environment-mediated interaction between foraging ants helps expedite the process of finding food and carrying it back home \cite{sumpter2003nonlinearity, jackson_longevity_2006,couzin2009collective,ants}. Inspired by the amazing ability of a cooperating collective to improve search efficiency by chemically imprinting the memory of past events onto its surroundings; we ask a fundamental question: can simple physical interactions between a single, non-intelligent, searcher and its environment  facilitate target location?

Here, we address this question within a controlled experimental setup: a bristle robot searching in a field of movable obstacles. Bristle robots are devices that convert the kinetic energy of vibration into forward propulsion using their flexible bristle-like legs  \cite{giomi_swarming_2013}. They exhibit persistent random motion \cite{giomi_swarming_2013,dauchot_dynamics_2019} and have been applied to study swarm robotics \cite{deblais2018a,boudet2021} and statistical mechanics out of thermal equilibrium \cite{baconnier2022, dasgupta2022non, Engbring2023}. 

In our experiment, collisions with the movable obstacles allow the bristle robot to imprint memory onto its environment and simultaneously ``sense'' the presence of trails it formed previously. The robot is given a fixed time window to explore the arena. At the end of this time interval, to mimic a home return via an existing trail, the robot is returned to its starting position without erasing the trails it created previously. This process is performed repeatedly, resulting in  significant reshaping of the arena such that it is partitioned into clear trails and denser obstacle islands. Comparing these experiments to ones in which we scramble the arena to erase the robot's footprints, we demonstrate the benefits of environmental memory on search with home returns. 

\subsection*{Encoding memory and sensing using mechanical interactions}\label{sec2}

Fig.~\ref{fig:setup}\textbf{A,B} depicts details of the experimental setup. We employ commercially available self-propelled bristle robots (bbots, Hexbugs nano-newton series) and analyze their motion in an arena (outer $150\times90$~$cm^2$ and inner $137\times63$~$cm^2$ dimensions) containing cubic mobile obstacles ($0.95$~$cm^3$ Perspex cubes, $m=1$~$gr$). Bbots are active particles that convert electric energy stored in a battery to vibrations. The tilted elastic legs of the bbots transform the vibrations to persistent forward motion (with some chirality,  Fig.~1). Since active particles tend to move along boundaries, we fashion the arena's edges in a way that injects the bbots inwards and away from the boundaries. The area fraction of the mobile obstacles in the arena was set to $\phi=0.3$.

When a bbot is introduced to the arena it clears obstacles from its path by colliding with them (see Figure~\ref{fig:setup}\textbf{C} and supplementary video S1), which is typical for active particles in crowded environments \cite{biswas2020first,dias2023environmental}. At later times, it can either return to  previously cleared trails, create new trails, or destroy existing ones. Thus, the trail pattern formed in the arena ``remembers" the bbot's path, which we will henceforth refer to as environmental memory or simply just memory. Similar environmental memory was observed on microscopic scales in a mixtures of active and passive particles \cite{dias2023environmental}. Since the bbot moves more freely when it revisits previously cleared trails, environmental memory also feeds back onto the bbot's motion.  We record the motion of the bbot using a webcam (BRIO 4K, Logitech) at a rate of 60 FPS. We use conventional particle tracking algorithms \cite{Crocker1996} to extract trajectories.


We perform resetting experiments using the following protocol; initially, the obstacles are scattered randomly across the arena. The experiments start by releasing a bbot from the center of the arena. The bbot then carves its path by pushing aside obstacles that it encounters. Following a fixed time interval $\tau=20~s$ from its release, the bbot is plucked out of the arena and repositioned at the origin. Thus, the bbot's position is reset (supplementary video S2). This resetting is aimed to mimic the return home via an existing trail. The bbot is positioned at a new orientation after each resetting event to ensure uniform sampling of the initial direction of motion. The manual resetting process is cut out of the recorded trajectories of the bbot. An excerpt of a typical trajectory, including two resetting events, is shown in Fig.~\ref{fig:setup}\textbf{D}.




\subsection*{Environmental memory enhances mobility}\label{sec3}


To study the effect of environmental memory on the steady-state distribution of the bbot and its passage times (PT) to certain locations in the arena, we perform two sets of experiments: with and without environmental memory.  In the experiments with environmental memory, only the bbot's position is reset while obstacles are untouched. In contrast, in the experiments without environmental memory, we manually rescatter the obstacles upon resetting. Rescattering of the obstacles results in full resetting of the system: the bbot's position and its environment. We use a resetting protocol in which the bbot is reset at constant time intervals, i.e., sharp resetting \cite{pal2016,Bhat_2016,FPTsholomi,eliazar2020mean}. In addition, in both experiment versions, we reset the bbot's position when it reaches the arena's boundaries. We stress that, regardless of the resetting event type, the sharp resetting timer starts fresh when the bbot's position is reset to the origin.


\begin{figure*} [t]
    \centering
    \includegraphics[width=\textwidth]{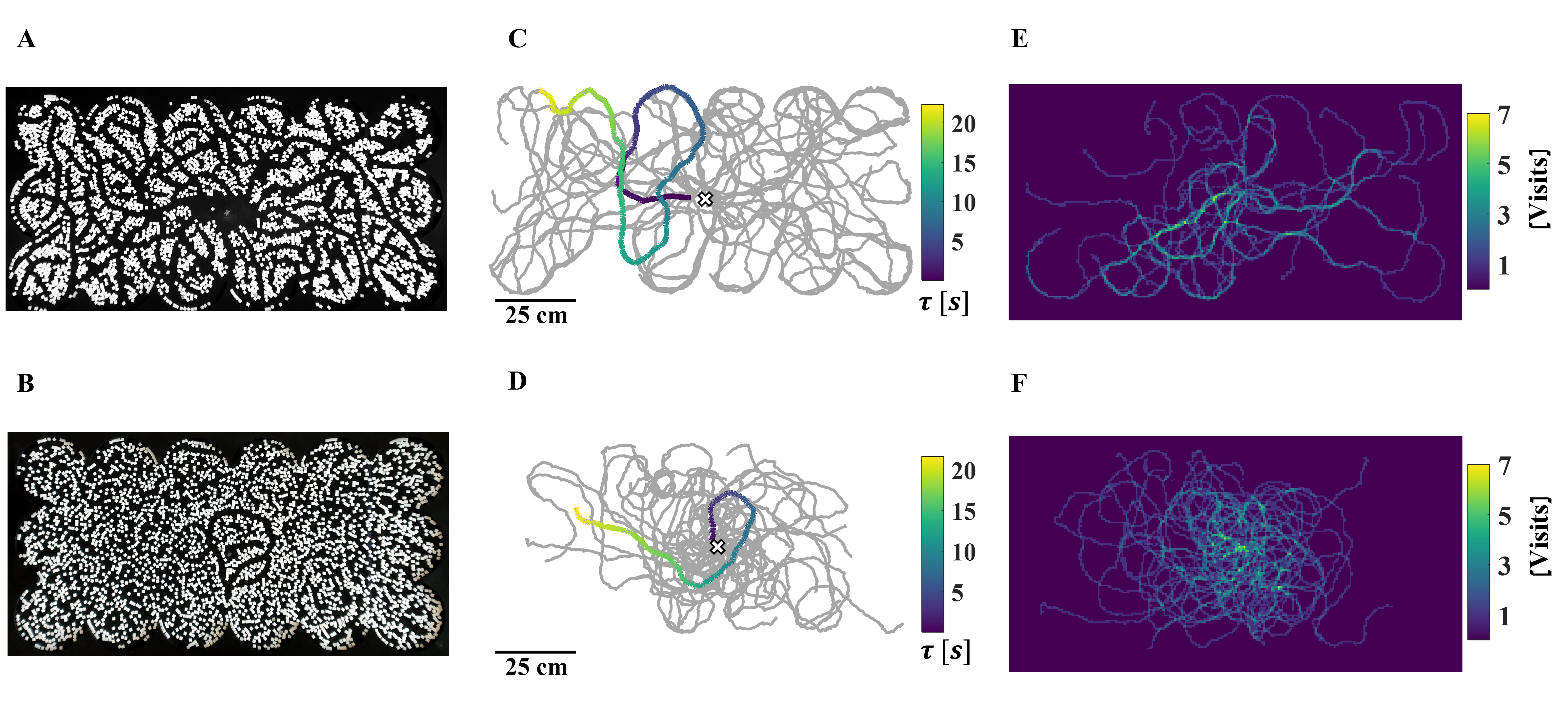}
    \caption {Comparison of typical arenas and the bbot's motion with and without environmental memory, top and bottom panels, respectively. \textbf{A} and \textbf{B}, Images of typical arenas with and without memory, $30$~$min$ into the experiment. \textbf{C} and \textbf{D}, Trajectories of the bbot with and without memory (in gray). The colored trajectories show the evolution in time of a single trail. A white cross marks the origin. \textbf{E} and \textbf{F}, Histograms of the number of visits to different locations in the arena with $0.25$~$cm^2$ sized bins.}
    \label{fig:memory}
\end{figure*}

In Fig.~\ref{fig:memory}\textbf{A} and \textbf{B}, we show snapshots of the arena $30$ minutes into the start of the experiment. The difference between panel A (with memory) and panel B (without) results in different characteristics of the ensemble of trajectories shown in Fig.~\ref{fig:memory}\textbf{C,D} (gray lines). Near the origin, the trajectories are sparser in the system with memory and denser in the system without. This observation implies that, in experiments with environmental memory, the bbot tends to revisit and stabilize existing trails. As a result, we observe striking differences in trajectory patterns. Specifically, the occurrence of nded trails is enhanced in the presence of environmental memory. We further visualize this effect in Fig.~\ref{fig:memory}\textbf{E,F}, where we present histograms of the number of visits to different locations in the arena. Quantifying the persistence length of the bbot we find $\ell_p=14.3~cm$ (with memory) and $\ell_p=10.0~cm$ (without), which provides quantitative support to the observations made above.

One of the hallmarks of random motion with resetting is the emergence of a non-equilibrium steady state for the position of the reset particle \cite{Evans2020}. Such a steady state is also reached in our experiments when considering the position of a bbot that is reset without keeping environmental memory. However, the situation is more complicated when we allow the environment to retain the memory of the bbot's motion, as the arena itself also evolves with time. In experiments conducted with environmental memory, the bbot constantly creates and destroys previous trails. As a result, obstacles are dynamically rearranged until the bbot's environment reaches a quasi-steady state (see SM Figs.~2 and 3 for  quantitative assessment). 

\begin{figure*} [t]
    \centering
      \includegraphics[width=\textwidth]{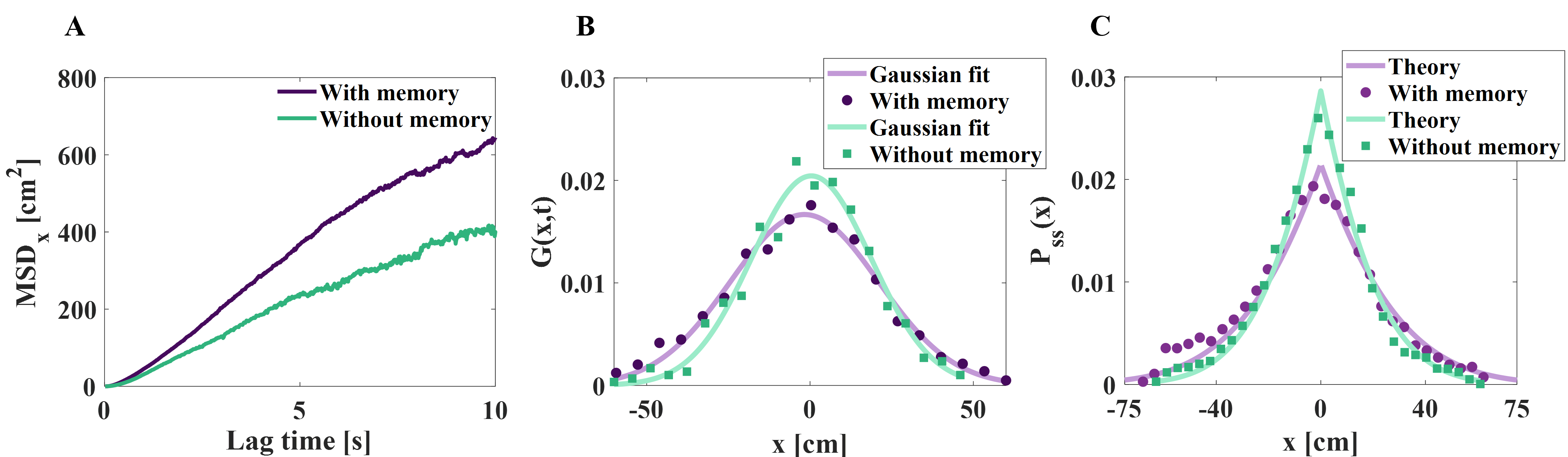}
    \caption{Characteristics of the bbot's motion within the quasi-steady-state phase, with and without memory. \textbf{A}, MSD as a function of the lag time, calculated from an ensemble of $N=1223$ (with memory) and $N=670$ (without memory) resetting events. \textbf{B}, Position distributions along the x-axis, at a lag time $t=5$~$s$ from a resetting event. Experimental measurements (markers) and fits to Eq.~\ref{Eq:propagator} (solid lines). \textbf{C}, Steady-state position distributions of the bbot along the x-axis, for resetting with and without memory. Experimental measurements (markers) and fits to Eq.~\ref{Eq:Pss} (solid lines). 
    In all panels, observe the enhanced mobility under environmental memory conditions.}
    \label{fig:msdxandss}
\end{figure*}

We characterize the bbot's motion and the resulting position distribution under these quasi-steady state conditions. First, we calculate the typical mean squared displacement (MSD) of the bbot along the long axis of the arena, which we henceforth denote as x. The MSD is computed by taking an average, over all $N$ resetting events, of the bbot's squared displacement 

\begin{equation}
    \langle\Delta x^2(t)\rangle=   \frac{1}{N} \sum_{i=1}^{N}(\Delta x_i(t))^2,
\end{equation}

\noindent where $\Delta x^2(t)=(x(t)-x(0))^2$, and $t$ is the lag time since the last resetting event.
In Fig.~\ref{fig:msdxandss}\textbf{A}, the MSD of the bbot is shown for the two sets of experiments, with and without memory. Both curves have a similar shape starting with a short super-diffusive segment that transitions into a diffusive (linear) regime, as expected for persistent motion in the presence of obstacles \cite{bechinger2016active}. Indeed, the bbot performs directed motion at short timescales and is scattered randomly by collisions with the obstacles at longer times. Eventually, the bbot reaches the boundaries and is reset. Therefore, we truncate the MSD curves at a cutoff that is smaller than the typical time it takes the bbot to reach the arena's boundaries. The diffusion coefficient of the bbot's motion is extracted from the slope of the MSD at the linear regime. We find $ D=34.7\pm0.3$~$cm^2/s$ and $ D=21.9\pm0.2$~$cm^2/s$ with and without memory, providing further support for the observation that the bbot's diffuses more efficiently with memory, i.e., when it can revisit existing trails.

To further quantify this effect, we measure the probability distribution of the bbot's position along the long axis of the arena  at a given time, $t$, indicating the lag time from the previous resetting event (Fig.~\ref{fig:msdxandss}\textbf{B}). In agreement with our previous observation, the distribution of position in a system with memory is wider. In addition, we find that this distribution is roughly Gaussian 
 
\begin{equation}
     G(x,t\vert x_0,0)=\frac{1}{\sqrt{4\pi D t}}e^{-\frac{(x-x_0)^2}{4D t}},
     \label{Eq:propagator}
\end{equation}

\noindent thus allowing us to map the problem to that of a particle performing free diffusion with an effective diffusion coefficient $D$, where $x_0$ is the origin.

The steady-state distribution of diffusion under Poissonian, i.e., constant rate, resetting has a cusp at the origin and exponentially decaying tails \cite{evans2011,Evans2020}. Here, resetting is conducted at fixed time intervals, but we observe a similar behavior of the steady-state position distribution $P_{ss}(x)$ of the bbot along the x-axis (Fig.~\ref{fig:msdxandss}\textbf{C}). Moreover, the functional form of $P_{ss}(x)$ is similar to that of a normally diffusing particle undergoing sharp resetting in an obstacle-free system, 
\begin{equation}
P_{\text{ss}}(x) =\frac{1}{\tau} \int_{0}^{\tau} \frac{1}{\sqrt{4\pi Dt}}e^{ -\frac{(x-x_0)^2}{4Dt} }  dt,
    \label{Eq:Pss}
\end{equation}
where we integrate $G(x,t\vert x_0,0)$  from Eq.~(\ref{Eq:propagator}) over the lag time. Integrating up to $\tau$, the length of the sharp resetting time interval, is sufficient since the evolution of our system at steady state statistically repeats itself in periods of $\tau$. This prediction of $P_{ss}(x)$ fits well the experimental data of both systems, with and without memory. The effective diffusion coefficients extracted from the fits are $D=34.4\pm4.3$~$cm^2/s$ and $D=19.4\pm1.7$~$cm^2/s$, with and without memory, respectively. The obtained values are similar to the ones extracted from MSD measurements. 


\subsection*{Environmental memory facilitates search}\label{sec4}

\begin{figure}
    \centering
    \includegraphics[width=0.5\textwidth]{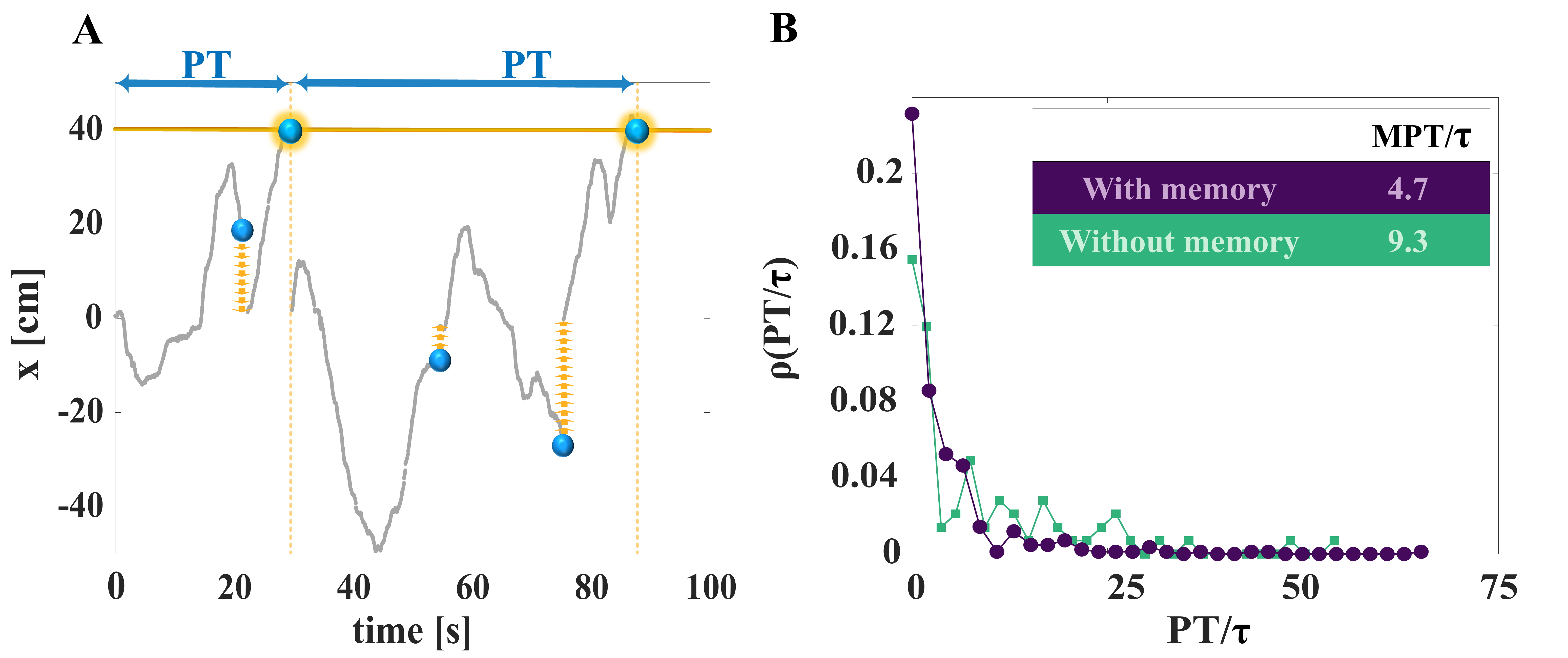}
    \caption{Characteristics of search under resetting. \textbf{A}, Projection of the bbot's trajectory onto the x-axis in a typical experiment showing two instances in which the bbot reached the target. The two PTs are marked using double-headed, blue arrows. The target is positioned at $x=40$~$cm$ and is marked  by a yellow line. Resetting events are marked using a series of arrows pointing back at the origin. \textbf{B}, Comparison between the PT probability distributions with and without memory. Here, the PT is taken in units of the time, $\tau$, between sharp resetting events. Markers come from experiments, and continuous lines are used as a guide to the eye. The MPTs with and without memory are indicated in the inset.}
    \label{fig:MFPT}
\end{figure}

Having established that memory enhances the bbot's mobility, we now turn to quantify the effect of memory on search.  We introduce a target line to our setup at $x=40$~$cm$  (see Fig~\ref{fig:MFPT}{\bf A}). We look at a sequential series of search processes, each starting at the origin and ending at the target. For each search process, we define the PT as the time it takes the bbot to reach the target starting from the origin. Note that the PT includes all the resetting events that happened prior to the bbot's arrival at the target. We also reset the bbot's position after it reaches the target.
In experiments without environmental memory, we manually rescatter the obstacles after each resetting event and when the bbot reaches the target.

\begin{figure*} [t]
    \centering
    \includegraphics[width=\textwidth]{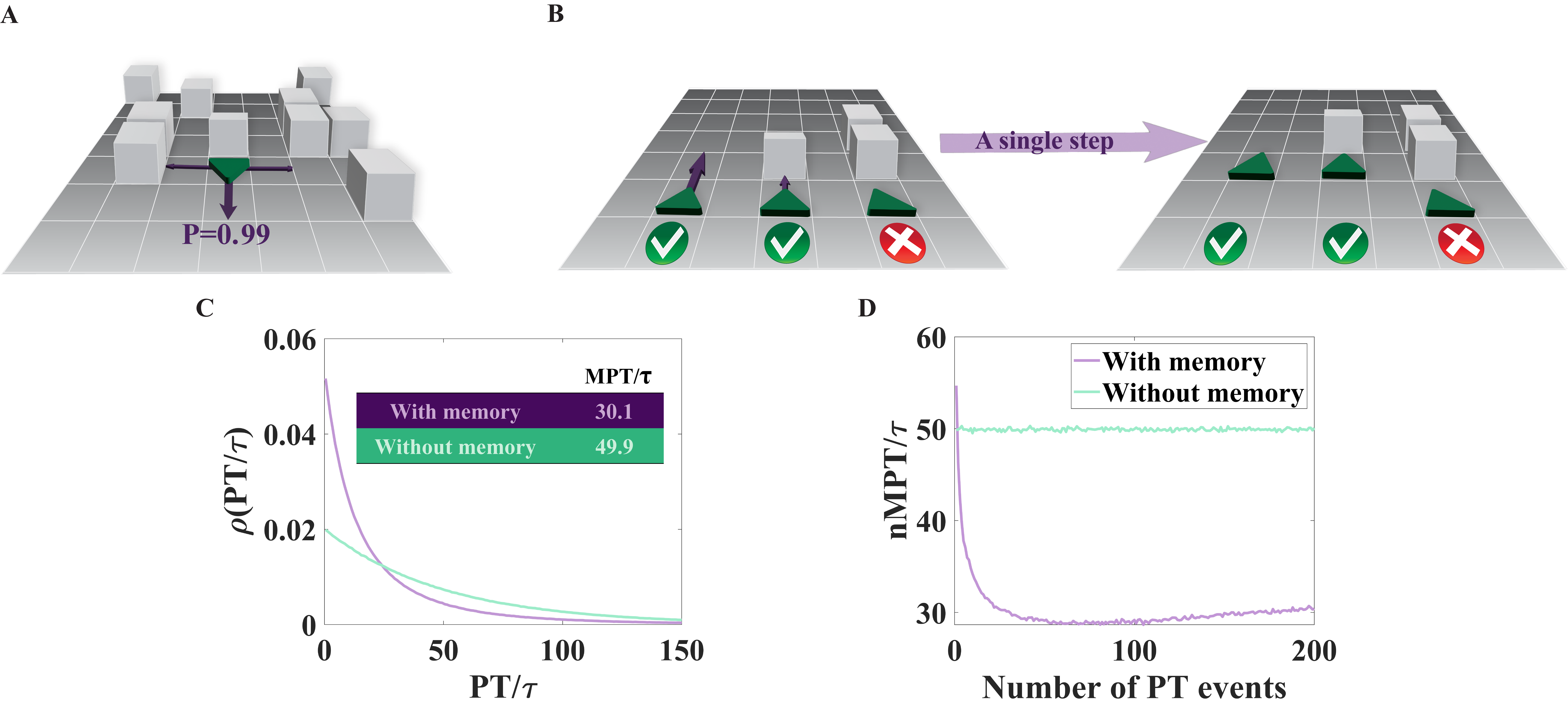}
    \caption{ \textbf{A} and  \textbf{B}, The persistent \textit{Sokoban} random walk and its laws of motion. \textbf{C}, Comparison between the PT probability distributions with and without memory. Here, the PT is taken in units of the time, $\tau$, between sharp resetting events.  The MPTs with and without memory are indicated in the inset. \textbf{D}, The mean n-th passage time (nMPT) as a function of the number of PTs since the beginning of the simulation. In simulations with memory, obstacles are scattered only at the beginning of the simulations. Hence memory is continuously encoded into the environment, and the nMPT decreases until it saturates at the steady state value reported in the inset of panel C. In contrast, in simulations without memory obstacles are re-scattered after every resetting event. Thus, the nMPT remains constant.}
    \label{fig:MFPTsim}
\end{figure*}


In some cases, the bbot does not reach the target during the entire 30~$min$ duration of the experiment. Therefore, we first measure the fraction of experiments in which the bbot reached the target. We find that for a system with memory, $N_{reach}/N_{total}=329/330\sim99\%$ whereas $N_{reach}/N_{total}=79/82\sim96\%$ for a system without memory. This indicates that the probability of reaching the target in systems with memory is slightly higher. 

For bbots that were able to reach the target, we also measure the distribution of PTs, $\rho(\cdot)$, and plot it for experiments with and without memory in  Fig.~\ref{fig:MFPT}{\bf B}. Averaging over these distributions, we  find that mean passage times (MPTs) are given by $\langle PT/\tau\rangle=4.7\pm0.4$ and $\langle PT/\tau\rangle=9.3\pm1.2$, for systems with and without memory respectively. Note, that these numbers report means and standard errors of the means in units of the resetting time. We thus find that memory significantly reduces the time it takes a bbot to reach the target line in our experiment. 

\subsection*{Revealing the build up of environmental memory}\label{sec4}

Memory in the environment is built gradually, increasing with every resetting event and the subsequent redistribution of obstacles in the arena. As a result, we also expect the mean passage time (MPT) to decrease gradually, as time progresses, from a value close to that of a system without memory to that of a system with memory. The latter is expected when the obstacles' configuration in the arena reaches a steady state. However, capturing this transition requires much more data than is realistically attainable in experiments. Thus, instead, we use Monte-Carlo simulations to show that the predicted effect is also seen in a stylized model that captures the essential physics governing our experiment. To this end, we adapt the \textit{Sokoban} random walk, which has recently been introduced to explore the effect of tracer-media interactions on motion and transport in disordered media \cite{bonomo2022sokoban}.

In the \textit{Sokoban} simulations, a random walker of unit size moves on a rectangular, 151$\times$91, arena with mobile obstacles of unit size scattered throughout (Fig.~\ref{fig:MFPTsim}{\bf A}). The area fraction of the mobile obstacles in the arena was set to $\phi=0.3$, as in the experiments. At each time step, the random walker tries to move one step in a random direction relative to the direction of its last step. To imitate a bbot's motion, we make the walker persist by taking the probability of forward motion to be 0.99, and the probability of moving either left or right to be 0.005. Additional rules were implemented to take into account interactions between the random walker and the obstacles scattered across the arena (Fig.~\ref{fig:MFPTsim}{\bf B}). Namely, a move always takes place if the walker tries to move into an unoccupied site. Otherwise, the walker may try to move into an occupied site by pushing an obstacle one site forward (in the direction of motion). However, such a move will only take place if the site to which the obstacle is pushed is vacant. In other words, unlike in experiments, the \textit{Sokoban} walker cannot push multiple obstacles simultaneously (supplementary video S3). If a forbidden step is attempted, the walker stays put at that given time unit. 

Next, we performed resetting simulations using a protocol that is almost identical to that used in the experiments. Obstacles are scattered randomly across the arena, and the simulation starts by releasing the walker at its center. The walker then carves its path by pushing aside obstacles that it encounters. As in the experiments, we implement sharp resetting by returning the walker to the origin at constant time intervals of $\tau=2000$ time steps. To ensure uniform sampling of the initial direction of motion, the walker is positioned at a new orientation after each resetting event.

We performed two sets of simulations: with and without environmental memory. In simulations conducted with environmental memory, the arena is left untouched, i.e., it is not randomized after resetting the walker's position. As a result, obstacles are dynamically rearranged until the environment reaches a steady state (SM Fig.~4A). In contrast to the experiment, where only a quasi-steady state is achieved, in the \textit{Sokoban} simulations,  a steady state is characterized by obstacle configurations that the walker can no longer modify. Hence, we eventually observe obstacle configurations that are fixed in time. Sample trajectories for simulations with and without memory are shown in SM Fig.~4B,C.

To characterize the effect of memory on the walker's motion, we repeated the analysis reported in Fig.~\ref{fig:msdxandss} for the simulated data. At short times ($t \leq 50$ steps), we observe the effect of directed motion on the MSD, per expectations and experimental observations. At longer times, the MSD becomes linear with a diffusion constant of $D=0.119\pm0.014$ and $D=0.073\pm0.002$ with and without memory, respectively (SM Fig.~5{\bf A}). Note that the forward motion probability of the walker was chosen such that the ratio of MSDs, with and without memory, will be similar to the experimentally measured one. As in the experiment, we find that the probability distribution of the walker’s position along the long axis of the arena is approximately Gaussian. The enhanced mobility of the walker under environmental memory conditions is clearly reflected on the propagator level (SM Fig.~5{\bf B}) and when comparing steady-state position distributions along the x-axis (SM Fig.~5{\bf C}).  

Having established that memory enhances the walker's mobility, similar to its observed effect in experiments, we turned to quantify the effect of memory on search. We introduced a target line to our simulations at $x=40$ and collected statistics for the PT to the target. In Fig.~\ref{fig:MFPTsim}{\bf C}, we compare the probability distributions of the PT for simulations conducted with and without memory. Following the experiments, simulations with memory were conducted at steady state conditions of the arena. Similar to the experiments, here too we find that a small fraction of walkers fail to reach the target. These are excluded from the PT distributions. Averaging over the PT distributions, we find that MPTs are given by $\langle PT/\tau\rangle=30.1$ and $\langle PT/\tau\rangle=49.9$, for systems with and without memory respectively. 

Finally, we characterize the change in the MPT due to the structural evolution of the environment as memory is built into the arena. We do so by calculating  the mean n-th passage time (nMPT), which is averaged over many realizations as a function of the number of PTs since the start of the experiment. In  simulations without memory, obstacles are re-scattered in the arena after each resetting event. We thus expect the nMPT to remain constant, i.e., not to depend on the passage number and the time elapsed since the onset of the simulation. In contrast, for simulations with memory, we expect the nMPT to decrease with time, evolving with the changes in the obstacles' distribution in the arena from a value close to that of a system without memory to that of a system with memory at steady state conditions. These expectations are in perfect agreement with simulation results that are plotted in Fig.~\ref{fig:MFPTsim}\textbf{D}. The transition observed in the simulations agrees with the two limiting conditions measured in experiments.

Concluding this section, we see that the broader implications of our experimental findings become evident through their manifestation in a different model system: the \textit{Sokoban} random walk. This stylized model was designed to broadly capture the fundamental physics underlying our experiment, but specific details of its microscopic dynamics are clearly different. Intriguingly, we find that the \textit{Sokoban's} motion must be persistent to reproduce the decrease in search time resulting from environmental memory. This finding emphasizes that environmental memory from the type studied here must be paired with persistent motion in order to benefit search.



\subsection*{Long-term memory leads to ergodicity breaking}\label{sec4}

Since the bbot effectively displays  diffusive behavior, it is only natural to ask if the measured values of the diffusion coefficients can be used to provide a theoretical estimate that captures the experimental MPTs. To answer this question, we construct a simplified model of our experiments. We assume that: (i) the bbot's motion can be described by  simple diffusion, (ii) motion along the x direction is independent of  motion along the y direction, and (iii)  resetting occurs after a fixed time $\tau$ or when one of the boundaries at the y direction is reached. This neglects  resetting events from the far boundary ($x=-68.5$~$cm$), which are relatively rare. For resetting without memory, these simplifications allow us to approximate the MPT using a known formula from the theory of first-passage under restart.

Letting $T$ denote the passage time without resetting and $R$ denote a statistically independent resetting time, we have $\langle PT\rangle=\frac{\langle min(T,R)\rangle}{\text{Pr}(T<R)}$ \cite{FPTsholomi}. To apply this formula in our case, we approximate $T$ with the analytically known first-passage time of diffusion in an interval with an absorbing boundary at $x=40$~$cm$ and a reflecting boundary at $x=-68.5$~$cm$. To approximate $R$, we take the minimum between the sharp resetting time $\tau$ and the analytically known first-passage  time of diffusion in an interval with absorbing boundaries at $y=+31.5$~$cm$ and $y=-31.5$~$cm$. The mean of the minimum of $T$ and $R$ can then be easily calculated, and the same goes for the probability that first-passage will occur before resetting. Overall, using $D=21.9$~$cm^2/s$ and $\tau=20$~$s$, this calculation gives $\langle PT/\tau\rangle\approx 6.04$. Performing a similar calculation for the  \textit{Sokoban} random walk without memory (note the slightly different arena dimensions), we plug in $D=0.073$ and $\tau=2000$, and find $\langle PT/\tau\rangle\approx 52$.

While the theoretical estimates show good agreement with \textit{Sokoban} data, there is a relative error of approximately $\approx35\%$ compared to the experimental results. One potential source of error is the non-diffusive nature of the bbot's motion at time scales shorter than the persistence time. To address this issue, we also conducted simulations of our simplified diffusion model, keeping the target line and resetting rules consistent with the experiment. For short time steps, $dt\ll\ell_p^2/2D$, we obtained an MPT that matched the theoretical prediction as expected. Increasing the simulation time step to $dt=\ell_p^2/2D \approx 2.28$~$s$, and repeating the  simulations we got $\langle PT/\tau\rangle \approx 8.16$, which agrees with experiments (to measurement error). Consequently, we conclude that the lion's share of the error associated with the theoretical prediction originates from the non-diffusive behavior of the bbot at short time scales.

Finally, we ask whether results obtained in the presence of environmental memory can also be understood in a similar fashion. At face value, the answer to this question should be no. The formula used above is based on the renewal assumption, which implies that the past is forgotten after a resetting event and thus passage attempts are statistically independent and identical. Clearly, this assumption does not hold for experiments with environmental memory. Surprisingly, when performing the same calculation as before to estimate the MPT theoretically with $D=34.7$~$ cm^2/s$ (the measured  diffusion coefficient with memory), we obtain $\langle PT/\tau\rangle \approx 3.54$. This value yields a relative error of approximately $\approx 25\%$, which is comparable to the error observed in the absence of memory. Furthermore, similar to the no memory case, this error almost vanishes when accounting for the non-diffusive behavior of the bbot at short time scales as before. 

We thus conclude that the memory in our experiment primarily influences the effective diffusion coefficient of the bbot; and that once this is known, correlations between passage attempts are not strong enough to create significant deviations from the renewal-based theoretical prediction. This can be understood by noting that  the bbot scrambles the arena during its motion, which renders correlations short-ranged in time.

The situation is drastically different for the  \textit{Sokoban} in the presence of environmental memory. Unlike the experiment,  \textit{Sokoban} arenas evolve until they reach a fixed configuration of obstacles that the random walker can no longer alter. This means that the renewal assumption, following resetting events, is asymptotically exact in the long time limit of a single trajectory. However, for $D=0.119$ obtained from \textit{Sokoban} simulations with memory, the theoretical estimate yields $\langle PT/\tau\rangle\approx 14.95$, which is half the true value. We attribute this discrepancy to the breakdown of ergodicity. The reported diffusion coefficient is based on an ensemble average over numerous arena configurations. In contrast, for each time trace, a single configuration becomes fixed. i.e., there is no averaging in time. To show that this is indeed the primary source of error, we modified the \textit{Sokoban} simulation such that a new \textit{steady-state configuration} of the arena is sampled with each resetting event. We then recover good agreement with the theoretical estimate, reducing the relative error to approximately $\approx 15\%$.

Concluding this section, we see that one significant difference between simulations of the \textit{Sokoban} random walk and our experiments is the time scale over which environmental memory persists. In the \textit{Sokoban} model, the distribution of obstacles across the arena reaches a steady state configuration that is frozen in time. Thus, the system does not self-average over time, and ergodicity is broken. This ergodicity breaking is different from the inherent ergodicity breaking of resetting processes \cite{Remi2023}.  In contrast, in the experiment, obstacles constantly move: trails are formed and destroyed such that a dynamic quasi-steady state is reached. Crucially, the bristle robot continues to scramble the arena in its motion, which renders memory and correlation times finite. We expect this to change in experiments with higher obstacle densities (SM Fig.~6), which give rise to longer-lasting memory and non-ergodic behavior. Clearly, a new theoretical framework to understand resetting in the presence of memory and strong temporal correlations will then be required. A more detailed characterization of the effect of obstacle density on search and mobility is left for future work.

\subsection*{Discussion}\label{sec5}

By itself, the ability to mark and sense the environment does not improve or detract from the efficiency of a search process. Yet organisms, such as ants and bacteria found ways to expedite the search for nutrients by use of chemical markings that relay information to the future self (and other searchers) via the environment. It is natural to ask, what is the level of complexity required of a searcher to implement such mechanisms to better its search?  Can a simple physical mechanism facilitate search by encoding environmental memory? 

We conducted the first controlled experiment to measure the impact of environmental memory on motion and target location. We showed that even a single, non-intelligent searcher---a bristle robot---can expedite its search by taking advantage of mechanical interactions that combine the ability to mark and perceive the environment. Using a resetting protocol, we demonstrated that environmental memory---manifested as obstacle-free trails---significantly improved the robot's mobility; consequently leading to a broader distribution of the bristle robot's location and reduced search times. 

Here we focused on elucidating the primary effects of environmental memory on search, leveraging the theoretical framework of resetting to quantify its ramifications. While we focused on a relatively simple setting, broad conclusions coming from our study are expected to apply more generally.  For example, our choice of the return method---instantaneous returns---allowed us to focus on the effect of environmental memory on the search phase. Yet, conclusions remain the same when generalizing to a realistic return model wherein the searcher retraces its path back to the origin. In this model, the return time is equivalent to the outbound search time by construction. In our experiments, the outbound search time is the same for systems with and without memory since it is given by the sharp resetting time. Consequently, the return times contribute equally in systems with and without memory, and our findings are unaffected by this specific choice of return mechanism. One can also explore how different return mechanisms affect search with environmental memory. This question is a natural extension of our present study, which we leave for future work. 

Our results suggest that any memory encoding (and sensing) method, such as mechanical, chemical, or optical, which enhances the speed or spread of the searching agent, will lead to lower search times. Moreover, since memory encoded in the environment can also be used by other searchers, our study extends to search involving multiple searchers originating from the same place, although we neglect direct interactions between the searchers. Enhanced mobility will also expedite the detection of targets with varying sizes and dimensions. 


Our findings also present a new theoretical challenge. The current theoretical framework aimed to predict first-passage under resetting is based on the renewal assumption. This assumption is clearly violated where environmental memory is long-ranged, which leads to significant prediction errors. A new theoretical framework is required to better understand resetting in the presence of memory and strong temporal correlations. 

In conclusion, we studied the effect of environmental memory on search with home returns and showcased its benefits in well-controlled laboratory experiments and numerical simulations. It is now essential to explore how complexity influences environment-assisted search strategies by taking a fundamental physics perspective. For instance, how do nonlinear effects arising from environmental reorganization impact search efficiency? And what is the effect of different strengths, types, and durations of trail markings? Introducing agent complexity, such as memory, sensing, communication, and computational capabilities---which are common in swarm robotics---is expected to further amplify search performance. Yet, a quantitative understanding of the added value brought by these elements is clearly missing. Progress in this direction may also shed new light on how living organisms use environmental memory to facilitate search.

\subsection*{Acknowledgments}\label{Acknow}
Shlomi Reuveni acknowledges support from the Israel Science Foundation (grant No. 394/19). This project has received funding from the European Research Council (ERC) under the European Union’s Horizon 2020 research and innovation programme (Grant agreement No. 947731). Yael Roichman acknowledges support from the Israel Science Foundation (grants No. 988/17 and 385/21) and from the European Research Council (ERC) under the European Union’s Horizon 2020 research and innovation programme (Grant agreement No. 101002392). \thanks{S.R. and Y.R.  contributed equally to this work.}

\newpage

\end{document}